\documentclass[preprint,aps,nofootinbib,showpacs]{revtex4}
\usepackage{graphicx}
\newcommand{\as}{\alpha_s}
\newcommand{\BUp}{``Bottom-Up'' }
\newcommand{\rarr}{\rightarrow}

\begin{document}

\title{Thermalization of a color glass condensate and review of the 
``Bottom-Up'' scenario}
\author{Andrej El$^1$
\footnote{E-mail: el@th.physik.uni-frankfurt.de},
Zhe Xu$^1
$\footnote{E-mail: xu@th.physik.uni-frankfurt.de} and 
Carsten Greiner$^1$
\footnote{E-mail: carsten.greiner@th.physik.uni-frankfurt.de}}
\affiliation{$^1$Institut f\"ur Theoretische Physik, Johann Wolfgang 
Goethe-Universit\"at Frankfurt, Max-von-Laue-Str.1, 
D-60438 Frankfurt am Main, Germany}
\date{\today}



\begin{abstract}
The thermalization of a longitudinally expanding color glass condensate
with  Bjorken boost invariant geometry is investigated within microscopical
parton cascade BAMPS. Our main focus lies on the detailed comparison of 
thermalization, observed in BAMPS with that suggested in the \BUp scenario.
We demonstrate that the tremendous production of soft gluons via $gg \to ggg$,
which is shown in 
the \BUp picture as the dominant process during the early preequilibration,
will not occur in heavy ion collisions at RHIC and LHC energies,
because the back reaction $ggg\to gg$ hinders the absolute particle
multiplication. Moreover, contrary to the \BUp scenario, soft and
hard gluons thermalize at the same time. The time scale of
thermal equilibration in BAMPS calculations is of order
$\as^{-2} (\ln \as)^{-2} Q_s^{-1}$.
After this time the gluon system exhibits nearly hydrodynamical behavior.
The  shear viscosity to entropy density ratio has a weak
dependence on $Q_s$ and lies close to the lower bound of
the AdS/CFT conjecture.
\end{abstract}

\maketitle

\section{Introduction}
\label{intro}
The success of employing simple ideal hydrodynamics \cite{H01} in describing
the large values of the elliptic flow $v_2$ measured in Au+Au collisions at
the Relativistic Heavy Ion Collider (RHIC) \cite{PHENIX,STAR} indicates
that thermal equilibration of the produced quark
gluon system occurs on a short time scale and that the equilibrium is maintained until
the hadronization. It is of great interest to understand what mechanisms drive
the system to equilibrium. While coherent quantum effects like color
instabilities \cite{instab} may play a role at the very early stage when
the system is super dense, perturbative QCD (pQCD) bremsstrahlung 
processes are essential for momentum isotropization of quark gluon
matter \cite{XU,XU07} when the matter becomes more dilute due to the strong
longitudinal expansion.

The importance of pQCD bremsstrahlung was first raised in the
\BUp scenario \cite{BMSS}, which describes the thermal equilibration
of a color glass condensate \cite{CGC,KV} characterized by a saturation scale
$Q_s$. The main idea of the \BUp scenario is that while the hard gluons
with transverse momenta of order $Q_s$ degrade as the condensate evolves in
space time, soft gluons with transverse momenta much smaller than $Q_s$ are
populated due to pQCD $gg\to ggg$ bremsstrahlung. Soft
gluon production dominates the early stage of equilibration and a strong
parametric enhancement of the soft gluon number has been predicted. Within a short
time scale the soft gluon number becomes comparable to the initial number
of hard gluons. As soon as the radiated soft gluons achieve thermal
equilibrium and build up a thermal bath, the hard ones begin to loose
their energy to the thermal bath and subsequently thermalize on a later time scale.
A parametric time scale for overall thermalization is given by
$\tau_{\rm th} \sim \as^{-13/5} Q_s^{-1}$ \cite{BMSS}.

Because color glass condensate (CGC) \cite{CGC} is proposed as a possible
initial state of the quark gluon matter produced in high energy heavy ion
collisions, its thermalization is a highly interesting topic.
For instance, the thermalization of an idealized form of color glass
condensate \cite{M2000} was studied by Bjoraker and Venugopalan by 
\cite{BV} solving the Landau transport equation. Serreau and Schiff also
investigated the same topic in \cite{SeSch01} where they used the relaxation
time approximation to simplify the collision term in the Boltzmann equation.
A conclusion, which can be drawn from the two studies, is that
pQCD $gg\to gg$ collisions are not sufficient to achieve  thermal
equilibrium. Hence, the inelastic $gg\leftrightarrow ggg$ processes
are needed as emphasized in the \BUp scenario. 

The role of the pQCD  $gg\leftrightarrow ggg$ processes for thermal equilibration employing the 
color glass condensate initial condition is investigated in the present work for the first time within 
a full 3+1 dimensional transport calculation using the parton
cascade BAMPS \cite{XUG06}. The bremsstrahlung processes indeed lead to
rapid thermalization. In this paper we focus on how thermalization
occurs in BAMPS calculations. In particular, we investigate whether 
the \BUp scenario is realized as the proper way to describe thermalization
of  CGC, which may be formed in heavy ion collisions at RHIC and LHC
energies. Note that the back reactions of the bremsstrahlung, i.e.,
$ggg\to gg$ processes, are consistently incorporated in BAMPS, whereas
they are missing in the \BUp scenario. This, which is shown in our results,
leads to a different thermalization picture from the one suggested by the \BUp scenario.

This paper is organized as follows. In section \ref{sec2} we present
the parton cascade BAMPS and the setup. A boundary condition is introduced
to mimic a one-dimensional (longitudinal) expansion with Bjorken boost
invariance. In section \ref{sec3} the detail on the CGC initial conditions
is given. The importance of $ggg\to gg$ processes, which hinders
the multiplication of soft gluons, is discussed in section \ref{sec4}.
The numerical results based on BAMPS calculations are shown in
section \ref{sec5}. We compare the thermalization observed in BAMPS with
the \BUp scenario, determine the time scale of thermal equilibration for
various saturation scales $Q_s$ and coupling constants $\as$ and extract the shear viscosity
to entropy density ratio, $\eta/s$. A conclusion is given in section \ref{con}.

\section{Parton Cascade BAMPS}
\label{sec2}
BAMPS is a microscopical transport model, which solves the Boltzmann equations
for on-shell quarks and gluons using Monte Carlo techniques. In the present work only gluons are considered.
The main feature of BAMPS is the implementation of $2\to N$ and $N\to 2$ processes
in a consistent manner, which is based on the stochastic interpretation
of interaction rates \cite{XU}. The interaction rates or the interaction
probabilities are calculated locally in space where the phase space density
of particles, $f(p,x)$, is extracted numerically. BAMPS subdivides space into
small cells, which are regarded as the local positions where interactions
may occur. The smaller the cells the more local interactions can be realized.
However, smaller cells contain fewer particles and thus lead to larger
uncertainties in the extraction of $f(p,x)$. Therefore, we adopt the test particle
method to amplify the (pseudo)particle density by a factor $N_{\rm test}$.
The cross sections have to be reduced by the same factor to obtain the
same physical mean free path \cite{XU}.

In order to make comparisons with the \BUp scenario we calculate the space time
evolution of gluons in a tube with a radius of $R=5$ fm. The transverse wall
of the tube serves as a boundary to mimic one-dimensional (longitudinal)
expansion. Gluons are simply reflected on the cylindrical wall. In
the transverse plan a static ''spider web'' like cell structure is considered:
The polar angle $\phi$ and the radial length squared $r^2$ are divided
equally within $[0,2\pi]$ and $[0,R^2]$, respectively. This division gives
the same transverse area for all cells. For the numerical calculations we 
set $\Delta \phi=\pi/4$ and $\Delta r^2=5 \ \rm fm^2$. 
Longitudinally, space is divided
in $\Delta z$ bins, which have the same width in the space time rapidity 
$\eta=\frac{1}{2} \ln ((t+z)/(t-z))$. $\Delta \eta=0.2$ is set to be
a constant for all $\Delta z$ bins. The initial gluons are put into rapidity interval $\left[-3; 3\right]$. Our setup is adequate to the assumption
of Bjorken boost invariance \cite{bjorken}, which is used in this study.

Gluon interactions included in BAMPS are elastic pQCD $gg\to gg$
scatterings as well as pQCD inspired bremsstrahlung
$gg\leftrightarrow ggg$. The differential cross sections and the effective
matrix elements are given by \cite{GB82,biro,W96}
\begin{eqnarray}
\label{cs22}
\frac{d\sigma^{gg\to gg}}{dq_{\perp}^2} &=&
\frac{9\pi\alpha_s^2}{(q_{\perp}^2+m_D^2)^2}\,,\\
\label{m23}
| {\cal M}_{gg \to ggg} |^2 &=&\frac{9 g^4}{2}
\frac{s^2}{({\bf q}_{\perp}^2+m_D^2)^2}\,
 \frac{12 g^2 {\bf q}_{\perp}^2}
{{\bf k}_{\perp}^2 [({\bf k}_{\perp}-{\bf q}_{\perp})^2+m_D^2]}\,
\Theta(k_{\perp}\Lambda_g-\cosh y)
\end{eqnarray}
where $g^2=4\pi\alpha_s$. ${\bf q}_{\perp}$ and
${\bf k}_{\perp}$ denote the perpendicular component of the momentum
transfer and of the radiated gluon momentum in the center-of-mass
frame of the collision, respectively. $y$ is the momentum rapidity of
the radiated gluon in the center-of-mass frame, and $\Lambda_g$ is the
gluon mean free path, which is calculated  self consistently \cite{XU}.
A discussion of the present idealistic implementation of the LPM effect is given in \cite{XU07}.

The interactions of the massless gluons are screened by a Debye mass
\begin{equation}
m_D^2=\pi \, d_G \, \as \int \frac{d^3p}{(2\pi)^3 p} N_c \,f_g
\end{equation}
where $d_G=16$ is the gluon degeneracy factor for $N_c=3$.
$m_D$ is calculated locally using the gluon density function $f(p,x)$
obtained from  BAMPS. The suppression of bremsstrahlung
due to the Landau-Pomeranchuk-Migdal effect is taken into account
within the Bethe-Heitler regime using a step function
in Eq. (\ref{m23}).

The interaction rates per particle are obtained \cite{XU} by
\begin{eqnarray}
\label{r22}
R_{22} &=& n \langle v_{\rm rel} \sigma_{22} \rangle_2 \,,\\
\label{r23}
R_{23} &=& n \langle v_{\rm rel} \sigma_{23} \rangle_2 \,,\\
\label{r32}
R_{32} &=& \frac{1}{2} n^2 \left \langle \frac{I_{32}}{8E_1E_2E_3} 
\right \rangle_3
\end{eqnarray}
for $gg\to gg$, $gg\to ggg$, and $ggg\to gg$, respectively,
where
\begin{eqnarray}
\sigma_{22} &=& \frac{1}{2!} \int_0^{s/4} dq_{\perp}^2 \,
\frac{d\sigma^{gg\to gg}}{dq_{\perp}^2} \,,\\
\sigma_{23} &=& \frac{1}{2s} \frac{1}{3!} 
\int d\Gamma^{\prime}_1 d\Gamma^{\prime}_2 d\Gamma^{\prime}_3 \,
|{\cal M}_{gg\to ggg}|^2 \, (2\pi)^4
\delta^{(4)}(p_1+p_2-p^{\prime}_1-p^{\prime}_2-p^{\prime}_3) \,,\\
\label{I32}
I_{32} &=& \frac{1}{2!} \int d\Gamma^{\prime}_1 d\Gamma^{\prime}_2\, 
|{\cal M}_{ggg\to gg}|^2
\, (2\pi)^4 \delta^{(4)}(p_1+p_2+p_3-p^{\prime}_1-p^{\prime}_2)
\end{eqnarray}
where $d\Gamma^{\prime}_i=d^3p^{\prime}_i/(2\pi)^3\,2E^{\prime}_i$,  
$|M_{ggg\to gg}|^2=|M_{gg\to ggg}|^2/d_G$, $s$ is the invariant mass
for the interaction, $v_{\rm rel}=s/2E_1E_2$ denotes the relative velocity
of two incoming gluons and $\langle \cdot \rangle_2$ and 
$\langle \cdot \rangle_3$ symbolize the average within ensembles
of incoming gluon pairs and triplets, respectively. For each gluon 
pair and triplet positioned in a cell unit with volume $\Delta V$
the transition probability within a time interval $\Delta t$ is given 
\cite{XU} by
\begin{eqnarray}
P_{22} &=& v_{\rm rel} \sigma_{22} \frac{\Delta t}{\Delta V}\,,\\
P_{23} &=& v_{\rm rel} \sigma_{23} \frac{\Delta t}{\Delta V}\,,\\
P_{32} &=& \frac{I_{32}}{8 E_1E_2E_3} \frac{\Delta t}{{\Delta V}^2}\,,
\end{eqnarray}
respectively, as derived directly from the transition rates
(\ref{r22})-(\ref{r32}). Note that $R_{32}=3R_{23}/2$ for thermal equilibrium.
The factor $3/2$ indicates the ratio of the number of identical
particles in the initial state of the $ggg\to gg$ and $gg\to ggg$ interaction.

\section{CGC Initial Conditions}
\label{sec3}
For the initial condition a gluon distribution of a color glass condensate \cite{CGC} is applied. 
The theory of a color glass condensate is given by the saturation picture, 
which assumes that the parton distribution in a hadron or nuclei saturates at high energies as a result of competition between QCD bremsstrahlung and annihilation processes. 

The CGC initial condition  used in our simulations  consists of gluons with $p_T<Q_s$, which are produced by the nonperturbative part of the  nucleus-nucleus interaction. $Q_s$ denotes the saturation momentum, which is the typical momentum of gluons in the CGC. It is close to $2~GeV$ at RHIC and is expected to be $4-6~GeV$ at LHC\cite{KNV}. 
The color glass condensate is a state with high parton occupation number where th transverse momenta reach up to $Q_s$, whereas the occupation number drops to 0 for transverse momenta much larger than $Q_s$. Initially, most gluons have transverse momenta close to $Q_s$, whereas the longitudinal momentum of gluons in the central rapidity bin is approximately zero.

For the initial gluon distribution of Color Glass Condensate we employ an idealized and boost-invariant form \cite{BV}
\begin{equation}
f(x,p)=\frac{c}{\as N_c}\frac{1}{\tau}\delta(y-\eta)\Theta(Q_s^2-p_T^2)\label{IC}
\end{equation}
We take $N_c=3$ for SU(3). The factor \textit{c} in (\ref{IC}) is the  ``parton liberation coefficient'' which accounts for the transformation of virtual partons in the initial state into  on-shell partons in the final state, as introduced in \cite{KN}. The value of \textit{c} used in \cite{BV} was calculated for a SU(2) gauge theory to be $c=1.3$ \cite{KV} \footnote{This value we used in our previous calculations \cite{ECX,Dipl}.}. New SU(3) gauge theory calculations yield a  value of $c\simeq 0.4$ \cite{KNV,Lappi}, which we employ for the following calculations.

The initial particle density in the CGC approach is given by \cite{BV,M2000}
\begin{equation}
\frac{1}{\pi R^2}\frac{dN}{d\eta}=c\frac{N_c^2-1}{4\pi^2\as N_c}Q_s^2\label{initial_n}
\end{equation}
For the application of the Boltzmann equation, we need the phase space density to be smaller than unity. If phase space density is high, Bose enhancement factors should be considered in the collision integrals, which is not done in BAMPS model.
 
The initial gluons are produced at eigentime $\tau\sim \frac{1}{Q_S}$ and the initial phase space density $f(x,p)$ from Eq.(\ref{IC}) is infinite due to the delta function $\delta (p_z)\sim \frac{1}{\Delta p_z}$. Later the distribution in longitudinal momentum space broadens due to $2 \to 2$ (or $2\to 3$) collisions and the occupation number becomes finite. Our cascade starts at time $\tau_0=\frac{c}{\as N_c}\tau_i$  where $\tau_i\cong{1 \over Q_s}$. At this time the parton distribution function in Eq.(\ref{IC}) is still larger than unity. However, the same initial time has been applied in \cite{BV}. In the \BUp picture at a time $\tau\sim \as^{-3/2} Q_s^{-1}$ the distribution should become less than 1. We note that we do not employ the Bose enhancement factor $(1+f)$ within the Boltzmann collision terms. Hence, as long as $f$ is larger than 1 we underestimate the rates.

In the following we present the results of simulations for $Q_s=2$, $3$ and $4~GeV$, i.e., energies relevant for RHIC and LHC.

\section{Soft gluon production and annihilation}
\label{sec4}
The basic assumption of the \BUp thermalization picture is that soft gluons are continually emitted due to inelastic $2\rarr 3$ bremsstrahlung processes,
which increase in the soft the gluon number and leads to formation of a thermal
bath. However, the annihilation processes, which are the back reactions
of bremsstrahlung and missing are in the \BUp scenario, will hinder
the soft gluon production due to detailed balance. The annihilation
processes become dominant, when the system is oversaturated. Using
the CGC initial conditions in Eq.(\ref{initial_n}) we estimate whether
a net production
of gluons is possible during the subsequent space time evolution.

\begin{figure}[ht]
\begin{center}
\includegraphics[height=7cm]{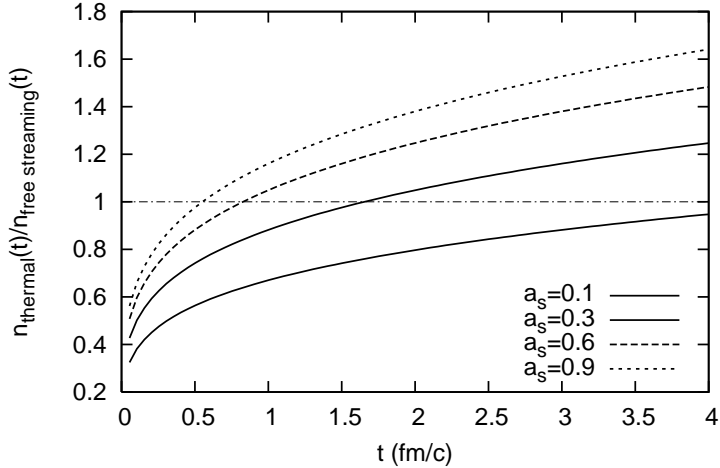}
\end{center}
\caption{
$n_{\rm th}(\tau_{\rm th})/n(\tau_{\rm th})$ ratio in Eq.
(\ref{ratio}) as a function of $\tau_{\rm th}$ for $Q_s=3$ GeV and
$\alpha_s=0.1$, $0.3$, $0.6$, and $0.9$, respectively.
}
\label{ratio_pic}
\end{figure}

Assuming a free streaming of CGC initial condition, the energy and particle densities
are given by
\begin{eqnarray}
e(\tau) &=& e(\tau_0)\, \frac{\tau_0}{\tau}=c\frac{N_c^2-1}{6\pi^2\as N_c}
Q_s^3\frac{1}{\tau} \,, \\
n(\tau) &=& n(\tau_0)\, \frac{\tau_0}{\tau}=c\frac{N_c^2-1}{4\pi^2\as N_c}
Q_s^2\frac{1}{\tau} \,.
\end{eqnarray}
Also assuming an instantaneous thermalization at time
$\tau_{\rm th}$ with
\begin{equation}
e(\tau_{\rm th})=e_{\rm th}(\tau_{\rm th})=
\frac{48}{\pi^2}T^4(\tau_{\rm th})\,,
\end{equation}
one obtains
\begin{equation}
n_{\rm th}(\tau_{\rm th})=\frac{16}{\pi^2}\,T^3(\tau_{\rm th})=
\frac{16}{\pi^2}\left(\frac{1}{288}\frac{N_c^2-1}{\as N_c}\, c \, 
Q_s^3\right)^{\frac{3}{4}} \left(\frac{1}{\tau_{th}}\right)^{\frac{3}{4}}\,.
\end{equation}
The ratio
\begin{equation}
\label{ratio}
\frac{n_{\rm th}(\tau_{\rm th})}{n(\tau_{\rm th})}=
24\left(\frac{1}{108}\right)^{\frac{3}{4}}\as^{\frac{1}{4}}c^{-\frac{1}{4}}
Q_s^{\frac{1}{4}}\tau_{\rm th}^{\frac{1}{4}}
\end{equation} 
estimates whether a net production or a net reduction of gluons will occur
at the early times of the expansion. If the ratio
$n_{\rm th}(\tau_{\rm th})/n(\tau_{\rm th})$ is larger than $1$, more
particles should be produced. Fig. \ref{ratio_pic} depicts 
the $n_{\rm th}(\tau_{\rm th})/n(\tau_{\rm th})$ ratio for fixed
$Q_s=3$ GeV and various values of $\as$ as a function of the thermalization time $\tau_{\rm th}$.
All curves start at values smaller than $1$, which indicates that for the chosen
parameters  indeed the annihilation processes would dominate the early
stage of equilibration. For an increase of the gluon number at early times,
as predicted in the \BUp scenario, the value of $\as$ has to be much larger
than $0.3$, or $Q_s$ has to be chosen much larger than would be given at RHIC
and LHC. For small coupling the particle number will start growing
if the time scale of thermalization is large.
In the full calculation of the Boltzmann equation the
gluon thermalization is more complicated than the simple consideration of
a free streaming and a subsequent instantaneous equilibration. However,
the behavior of the $n_{\rm th}(\tau)/n(\tau)$ ratio holds, as we will
shortly see.

\begin{figure}[ht]
\begin{center}
\includegraphics[height=7cm]{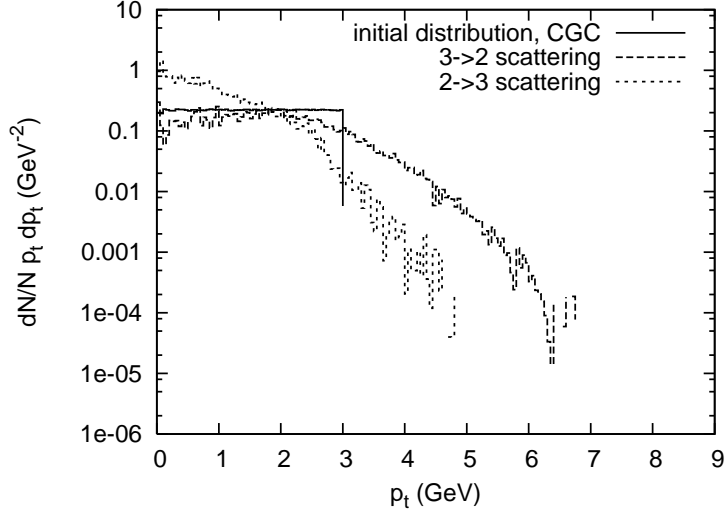}
\end{center}
\caption{Transverse momentum distribution after one single inelastic
scattering. Initial condition is a CGC with $\as=0.3$ and $Q_s=3$ GeV.
}
\label{single_scatt}
\end{figure}

The initial distribution of gluons is highly anisotropic in momentum space.
Most gluons have a transverse momentum of order $p_t\sim Q_s$. Populations
of the low (high) momentum gluons should be dominated by $2\to 3$ ($3\to 2$)
processes. Figure \ref{single_scatt} shows the gluon $p_t$ spectra after
one single $2\to 3$ or $3\to 2$ interaction.
While $3 \rightarrow 2$ processes
increase gluon number in high momenta, $2\rightarrow 3$
collisions lead to an enhancement of gluon number in soft momenta, which
resembles the \BUp scenario.

\section{Results: Thermalization of a CGC}
\label{sec5}
\subsection{Gluon number}
\label{sec5_1}
\begin{figure}[ht]
\begin{center}
\includegraphics[height=7cm]{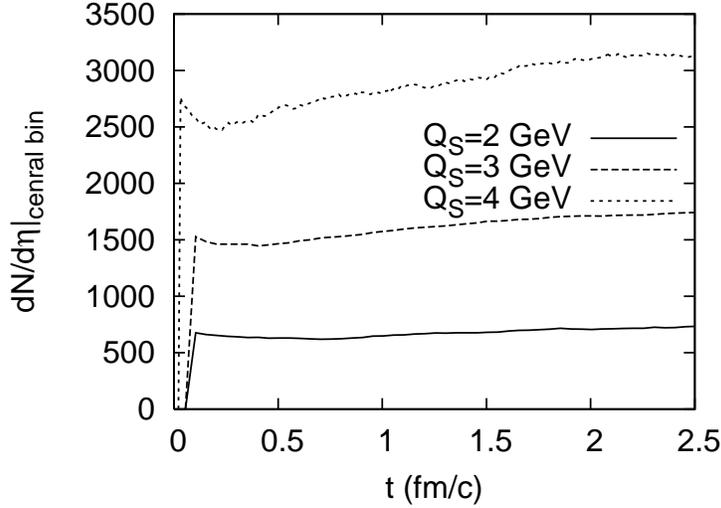}
\end{center}
\caption{
$dN/d\eta$, at midrapidity
($\eta \in [ -0.1,0.1]$) as a function of time. Results are obtained
using BAMPS for the initial CGC with $\as=0.3$ and $Q_s=2$, $3$, and 
$4$ GeV, respectively.
}
\label{dNdeta}
\end{figure}

\begin{figure}[ht]
\begin{center}
\includegraphics[height=7cm]{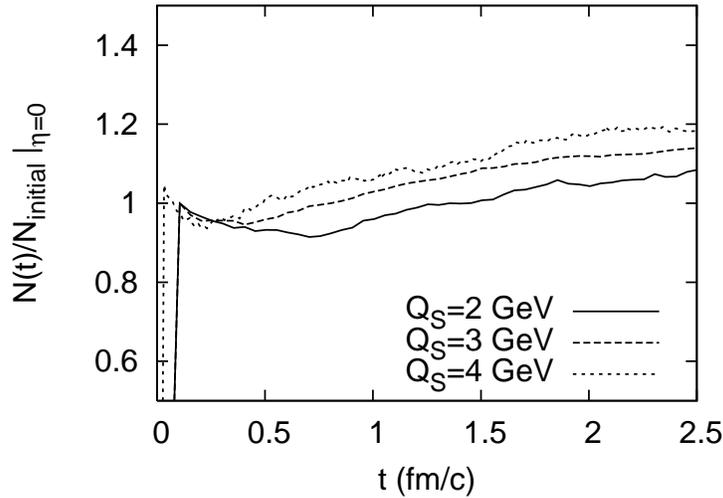}
\end{center}
\caption{
Ratio of the gluon number to the initial one in the central
space time rapidity bin.
}
\label{normdNdeta}
\end{figure}

\begin{figure}[ht]
\begin{center}
\includegraphics[height=4.8cm]{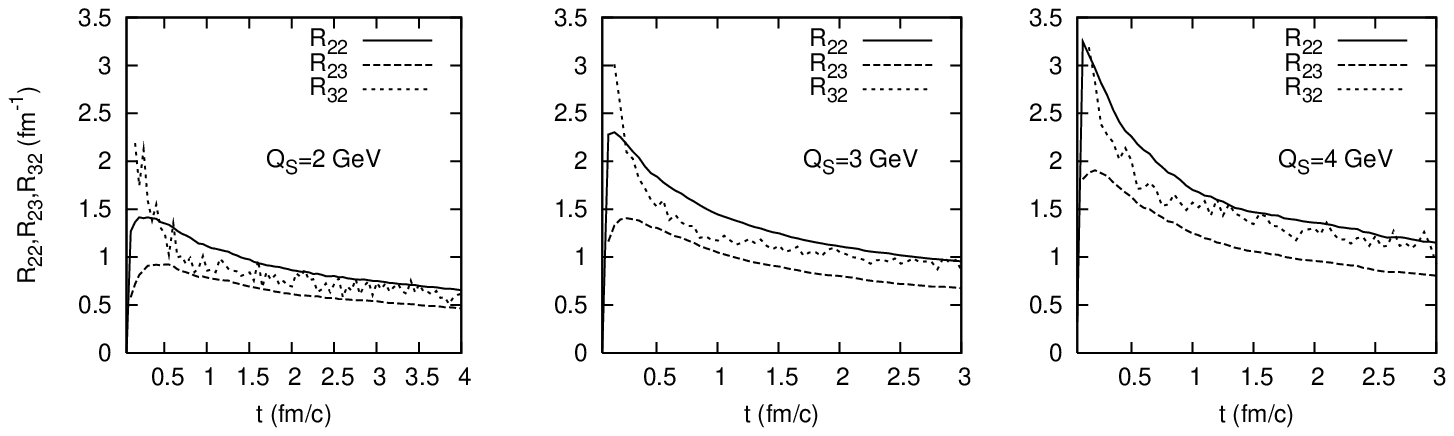}
\end{center}
\caption{
Interaction rates in the central space time rapidity bin, obtained from
BAMPS calculations for CGC with $\as=0.3$ and $Q_s=2$, $3$, and $4$ GeV.
}
\label{collrates}
\end{figure}

Fig. \ref{dNdeta} shows the gluon multiplicities per space time rapidity,
$dN/d\eta$, at midrapidity ($\eta \in [ -0.1,0.1]$) as function of time,
which are obtained using BAMPS with the  CGC initial conditions with $\as=0.3$ and
$Q_s=2$, $3$, and $4$ GeV, respectively.
Their ratios to the initial gluon number are depicted in
Fig. \ref{normdNdeta}.
The way that thermalization proceeds within the parton cascade calculations
does not resemble the way that has been advocated in the \BUp
scenario \cite{BMSS}. The strong parametric enhancement of the total 
gluon number at early times, as predicted by \BUp scenario,
is not observed in the cascade calculations. Instead, gluon annihilation 
occurs during the first $0.3-0.75$ fm/c for $Q_s=2-4$ GeV. This is
clearly due to the $3\to 2$ annihilation processes and indicates that the initial CGC
is oversaturated for the chosen values of $\as$ and $Q_s$. 
Figure \ref{collrates} shows the interaction rate of both elastic and
inelastic processes in the central space time rapidity bin.
The rate of $3\rightarrow 2$ processes is initially significantly higher than that
of $2\rightarrow 3$ processes, which leads to a decrease in the net
gluon number at very early times.

The gluon number begins to increase after $t\sim 0.3-0.75$ fm/c
(depending on the value of $Q_s$) when the system is close to kinetic
equilibrium and a quasi-hydrodynamical cooling sets in. Slow increase
in gluon number at late times is consistent with the collision rates,
$R_{23} \stackrel{>}{\sim} 2 R_{32}/3$, shown in Fig. \ref{collrates}.
(In chemical equilibrium $R_{23}=2 R_{32}/3$.) Assuming parton-hadron
duality, the final gluon multiplicities, $dN/d\eta \approx 700-1700$ for 
$Q_s=2-3$ GeV, are equivalent to $dN_{\rm ch}/d\eta \approx 470-1100$ for the 
total charged mesons, which are comparable with the RHIC data for
Au+Au most central collisions at 
$\sqrt{s}=130-200$ AGeV \cite{Adler,phobos}. 

\begin{figure}[ht]
\begin{center}
\includegraphics[height=7.5cm]{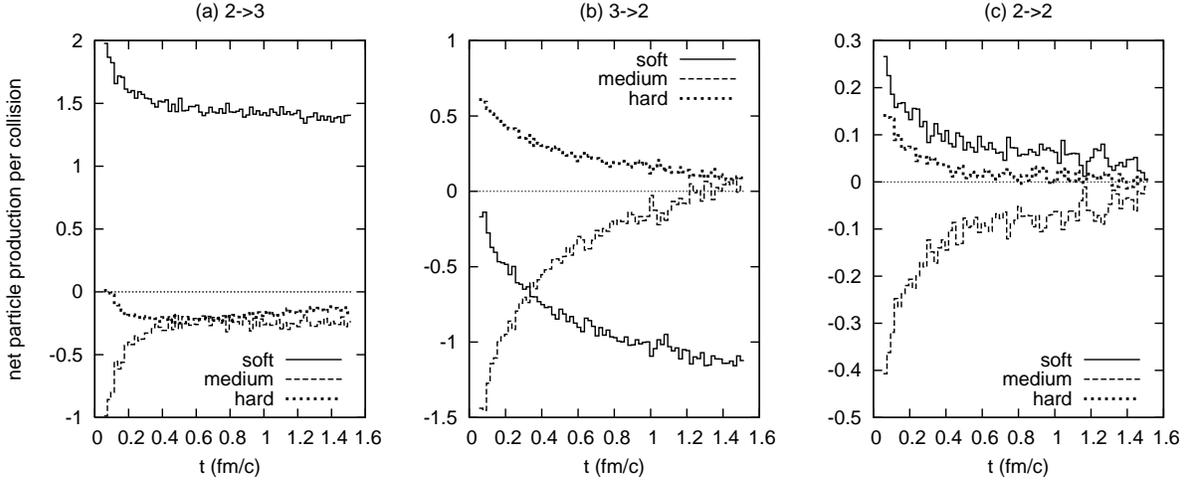}
\end{center}
\caption{Net production of soft, medium, and hard gluons in various
collisions. Results are obtained in the central space time rapidity bin
from a calculation performed for a CGC with $\as=0.3$ and
$Q_s=3$ GeV.
}
\label{coll_dist}
\end{figure}

As demonstrated in Fig. \ref{single_scatt}, $2\to 3$ collisions
lead to an enhancement in the number of the soft gluons, whereas $3\rightarrow 2$
processes initially increase number of  gluons with momenta higher than the saturation momentum
$Q_s$. We now study the changes in gluon momenta for various interaction channels
included in the calculations. For this we define a soft
momentum scale, which is $p_{\rm soft}^2 \le \as Q_s^2$ in the \BUp scenario,
to be $p_{\rm soft}=1.5$ GeV and a hard scale $p_{\rm hard}=Q_s$.
The medium gluons are denoted as gluons with 
$p_{\rm soft} < p_t < p_{\rm hard}$.
These definitions are in particular reasonable at early times since the
longitudinal momenta are very small. They are different from the definitions in \cite{BMSS}, where the authors set all the initial gluons to be hard. Fig. \ref{coll_dist} shows the
net gluon production of each type in $2\to 3$, $3\to 2$, and $2\to 2$
processes as function of time. 
The results are in terms of the difference between the number of outgoing and incoming
gluons of each type divided by the number of collisions. We see that
$2\to 3$ collisions increase the soft gluon number with a loss of
medium and hard gluons [see Fig. \ref{coll_dist}(a)], whereas $3\to 2$
processes increase the hard gluon number with the loss of soft and
medium gluons [see Fig. \ref{coll_dist}(b)]. Equivalently,
$2\to 3$ processes transfer energy from the medium and hard to the soft
sector, whereas $3\to 2$ processes transfer energy from the soft and
medium to the hard sector. Compared to $2\leftrightarrow 3$ processes,
$2\to 2$ convert few medium gluons into soft and hard ones [see Fig. \ref{coll_dist}(c)].
Common to all three collision types in Fig. \ref{coll_dist} is a continuous
degradation of the medium sector with the simplified CGC initial conditions.
As the system evolves towards equilibrium, energy is transferred from the 
medium into both the soft and hard sector.

\begin{figure}[ht]
\begin{center}
\includegraphics[height=7cm]{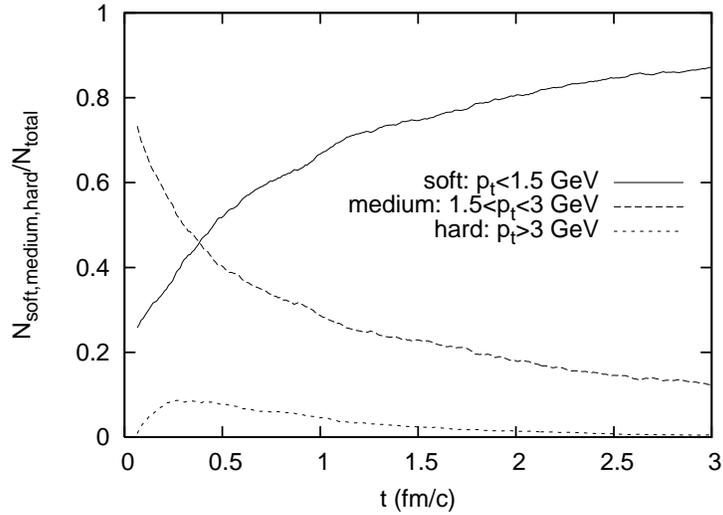}
\end{center}
\caption{Ratio of the numbers of the soft, medium and hard gluons to the
total number.
}
\label{percentage}
\end{figure}

The ratio of the numbers of the soft, medium and hard gluons to the total
number of gluons is depicted in Fig. \ref{percentage}.
The total gluon number is dominated by the medium sector until
$0.5$ fm/c and then by the soft sector after $\sim 1$ fm/c. 
Contrary to the \BUp picture, Fig. \ref{percentage} shows that the soft gluon number increases over a long period of time at the cost of the primary ``medium'' gluons. To repeat, the production
of soft gluons is effectively hindered by $3\to 2$ processes and, thus, cannot
exhibit a huge increase as predicted in the \BUp scenario.

\subsection{Kinetic equilibration and momentum isotropization}
\label{sec5_2}
\begin{figure}[ht]
\begin{center}
\includegraphics[height=7cm]{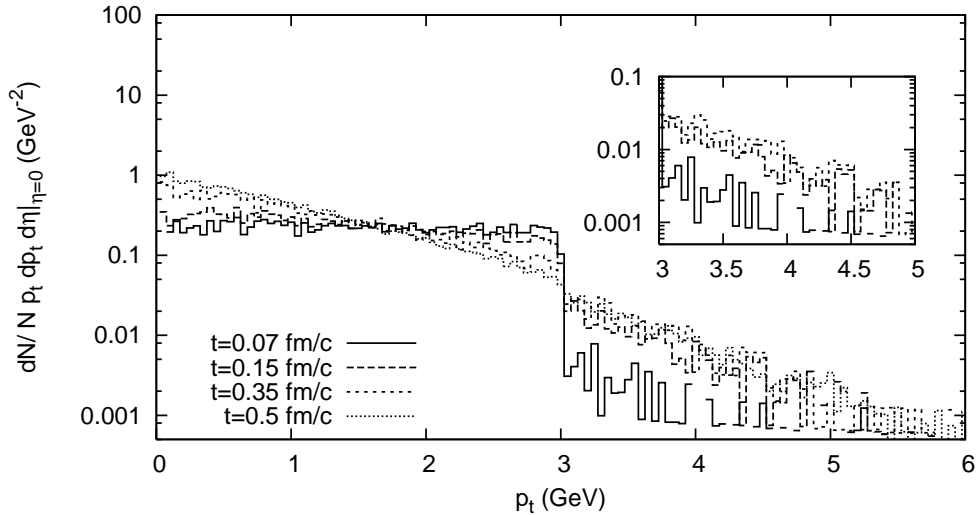}
\end{center}
\caption{Transverse momentum spectra in the central space time rapidity
and at various early times. The initial condition for the BAMPS calculation
is a CGC with $\as=0.3$ and $Q_s=3$ GeV.
}
\label{nnormpt_early}
\end{figure}

Fig. \ref{nnormpt_early} gives the transverse momentum spectra
in the central space time rapidity and at various early times (up to
$0.5$ fm/c) obtained from BAMPS for initial CGC
with $\as=0.3$ and $Q_s=3$ GeV.
After the expansion starts, energy flows immediately into both the soft
($p_t < p_{\rm soft}=1.5$ GeV) and hard momentum region
($p_t > Q_s$) where the populations rapidly increase, as also seen
in Fig. \ref{percentage}. Note that at $0.5$ fm/c the number of
soft gluons is of the same order as the number of harder gluons.

From Fig. \ref{nnormpt_early} we observe that the spectrum of high momentum
gluons achieves an exponential shape on a short time scale and almost as quick as
the soft gluons. However, they have different slopes. At $t=0.5$ fm/c
the entire spectrum is to a good extent in agreement with a thermal
fit taking
$T\simeq 0.67$ GeV, which is indeed very close to the effective temperature
of the system at this time [$T_{\rm eff}(t=0.5 \ {\rm fm/c})=E/3N=0.6$ GeV].
The transverse momentum spectrum achieves a thermal shape in hard
and soft regions almost simultaneously. The presence of a thermal bath 
of soft gluons seems not to be a necessary condition for the equilibration
of hard gluons. Again, this is different from the picture invoked in
the \BUp scenario.

In \cite{DNS} where the dynamics of SU(2) gauge fields in presence of
an initial anisotropy in momentum space is studied, it is shown that
the energy obtained from the particles by a Weibel-like plasma instability
does not lead to an exponential buildup of transverse magnetic fields. Rather
it is transferred into the ultraviolet modes via a rapid ``avalanche''.
This phenomenon, which was also discussed in \cite{AM}, is similar to
what we have observed during the very early thermalization of CGC (see 
Fig. \ref{nnormpt_early}). A more detailed study of such a highly nonlinear
phenomenon is certainly needed.

\begin{figure}[ht]
\begin{center}
\includegraphics[height=7cm]{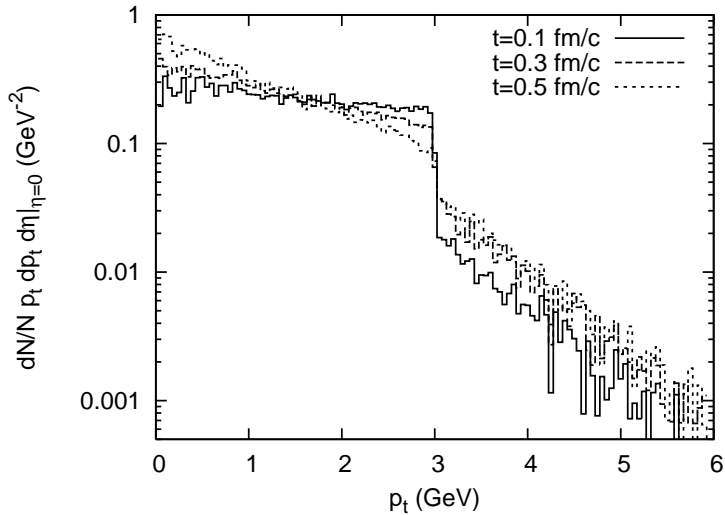}
\end{center}
\caption{Same as Fig.\ref{nnormpt_early}. The initial condition
is CGC with $\as=0.1$ and $Q_s=3$ GeV.
}
\label{pt_as01}
\end{figure}

For  CGC initial condition with $\as=0.1$ and $Q_s=3$ GeV the rapid 
''avalanche'' is again seen in from Fig. \ref{pt_as01}.
The number of hard gluons with an exponential shape increases
on the same time scale as in the case for $\as=0.3$.
This can be understood from the following consideration: Because at early
times $I_{32}$ [see Eq. (\ref{I32})] is roughly proportional to $\as^2$
and the gluon density is inversely proportional to $\as$, the initial
interaction
rate, $R_{32} \sim n^2 I_{32}$ [see Eq. (\ref{r32})], is approximately
independent of $\as$.

\begin{figure}[ht]
\begin{center}
\includegraphics[height=7cm]{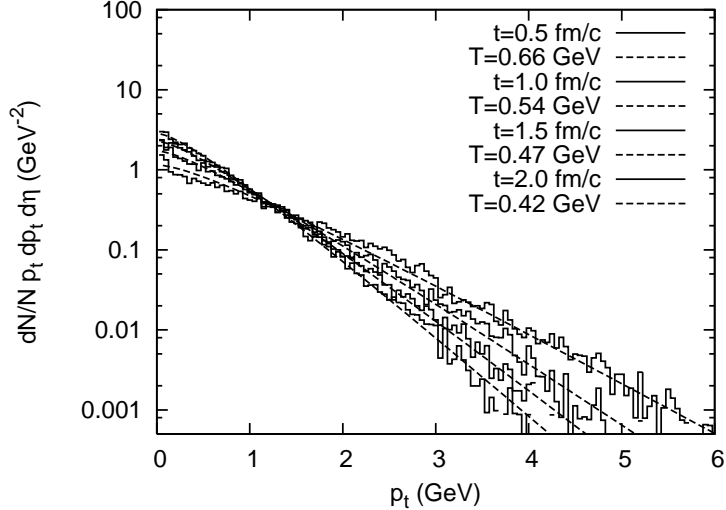}
\end{center}
\caption{Same as Fig.\ref{nnormpt_early} for later times.
}
\label{ptsp_long}
\end{figure}

Fig. \ref{ptsp_long} shows the transverse momentum spectra
for $\as=0.3$ at times larger than $0.5$ fm/c.
The spectra are compared with thermal fits using temperature $T$ as
a parameter.  The effective temperatures $T=E/3N$, which is extracted from the
simulations at various times, are found to be indeed close to the values chosen
for the fits. Thus, the transverse spectrum at $t=0.5$ fm/c looks almost thermal.
Moreover, the cooling of the system sets in, which makes the exponential
spectra steeper at later times. This behavior is characteristic for 
a hydrodynamical expansion.

Summarizing the results above, the thermalization of a CGC, studied
in the parton cascade BAMPS, is characterized by the following facts
to be opposed the \BUp scenario:
\begin{itemize}
\item Emission of hard ($p_t>Q_s$) gluons due to $3\rightarrow 2$
processes clearly dominates the very early evolution.
\item No strong enhancement of total gluon number is observed. The total
gluon number decreases slightly, until the system is nearly thermalized.
\item No thermal bath of soft gluons is built up at very early times.
\item Transverse momentum spectra achieves a thermal shape in the hard and
soft regions almost simultaneously.
\end{itemize}

\subsection{Time scale of thermalization}
\label{sec5_3}
\begin{figure}[ht]
\begin{center}
\includegraphics[height=6.5cm]{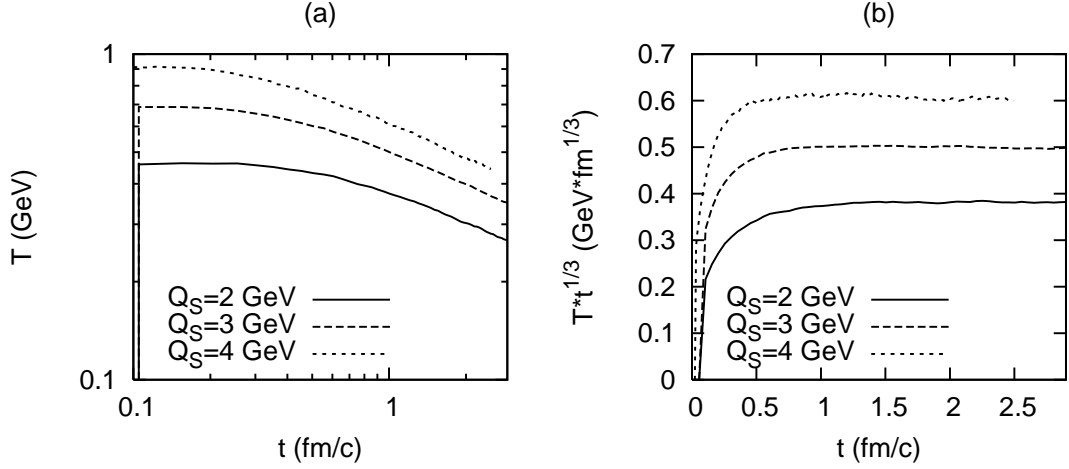}
\end{center}
\caption{(a) Effective temperature and (b) Scaled effective temperature.
}
\label{temp}
\end{figure}

\begin{figure}[ht]
\begin{center}
\includegraphics[height=8cm]{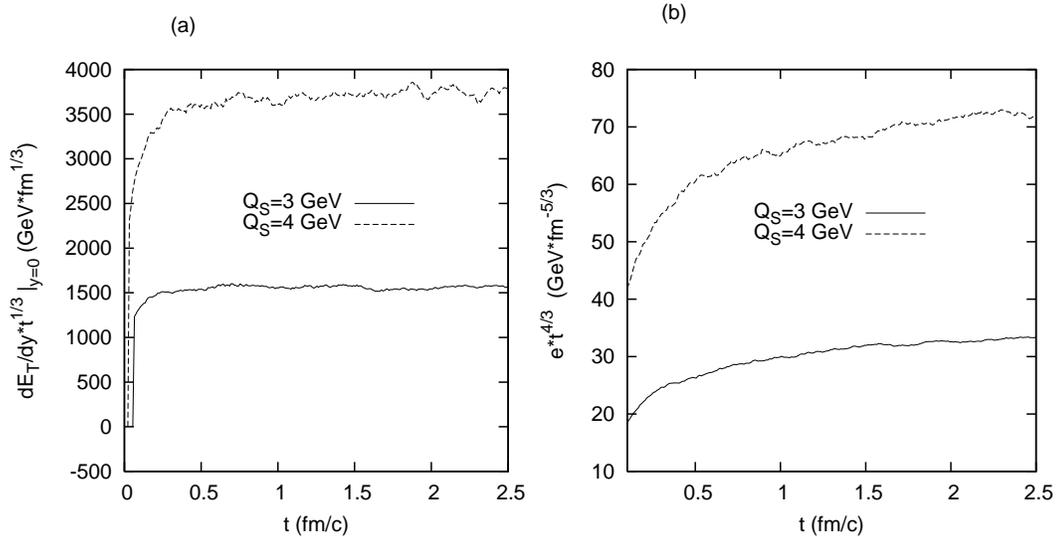}
\end{center}
\caption{(a) Scaled transverse energy per rapidity and (b)
scaled energy density.
}
\label{dEtdy}
\end{figure}

Next we extract the time scale when the system is more or less thermalized.
A parametric dependence of the time scale, $\sim \as^{-13/5} Q_s^{-1}$,
was given in the \BUp scenario \cite{BMSS}. Fig. \ref{temp} shows
the effective temperature, $T$, and the scaled one, $T t^{1/3}$, as a function
of time obtained with $\as=0.3$ and $Q_s=2$, $3$,
and $4$ GeV, respectively.
The effective temperature increases slightly at very early times
due to gluon annihilation and then falls approximately with $t^{-1/3}$
[see Fig. \ref{temp}(b)]. The same scaling is also found for the
transverse energy per rapidity at late times, as seen in Fig. \ref{dEtdy}(a),
where $dE_T/dy \cdot t^{1/3}$ is depicted.
The behavior at late times corresponds to a one-dimensional ideal
hydrodynamical expansion. Thus, we determine the time scale
of thermalization as the time, at which $T t^{1/3}$ becomes a constant.
The times extracted from Fig. \ref{temp}(b) read: 
\begin{equation}
\label{tau_th}
t_{\rm th}(Q_s=2\ {\rm GeV})=1.2\ {\rm fm/c},\ 
t_{\rm th}(Q_s=3\ {\rm GeV})=0.75\ {\rm fm/c},\
t_{\rm th}(Q_s=4\ {\rm GeV})=0.55\ {\rm fm/c}\,.
\end{equation}
The value for $Q_s=3$ GeV is consistent with the time scale ($0.5$ fm/c) 
at which the transverse momentum spectrum becomes exponential (see 
Fig.\ref{ptsp_long}). In addition, the values of $t_{\rm th}\cdot Q_s$ are
almost equal for fixed $\as$ and various $Q_s$, which verifies the
relation $t_{\rm th}\sim Q_s^{-1}$ as predicted in the \BUp scenario.

Although the expansion at late times is described nearly by ideal 
hydrodynamics, the collision rates are not infinitely high (see 
Fig. \ref{collrates}). Thus, the viscosity might be small but still nonzero
(as shown in the next subsection). The effect of the nonzero viscosity
is actually apparent in the scaled energy density, $e \,t^{4/3}$, depicted
in Fig. \ref{dEtdy}(b). Although $e \,t^{4/3}$ is constant in time according
to one-dimensional ideal hydrodynamics, $e \,t^{4/3}$ 
still increases at later times in the simulations.

\begin{figure}[ht]
\begin{center}
\includegraphics[height=7cm]{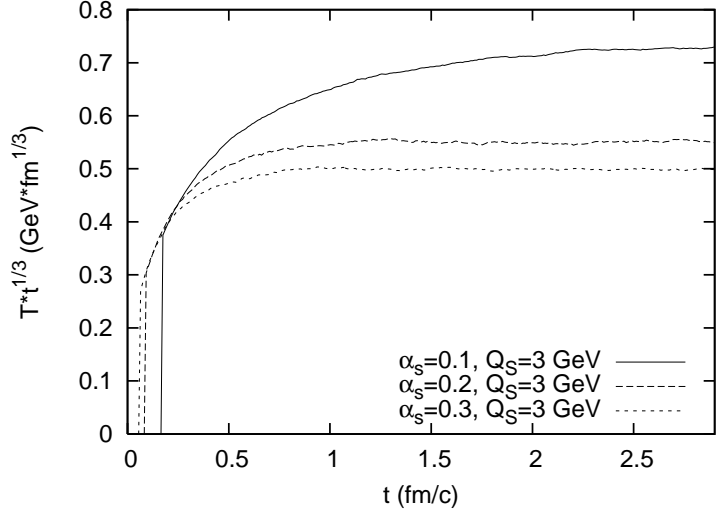}
\end{center}
\caption{Scaled effective temperature in simulations
with fixed $Q_s=3$ GeV and $\as=0.1$, $0.2$, and $0.3$.
}
\label{Ttas}
\end{figure}

For fixed $Q_s=3$ GeV and various $\as$ the time scale of thermalization
is extracted from Fig. \ref{Ttas} such that
\begin{equation}
t_{\rm th}(\as=0.1)=1.75\ {\rm fm/c},\
t_{\rm th}(\as=0.2)=1 \ {\rm fm/c}, \
t_{\rm th}(\as=0.3)=0.75 \ {\rm fm/c},
\end{equation}
and
\begin{equation}
t_{\rm th}\as^{13/5}=0.0044\ {\rm fm/c}(\as=0.1), \ 0.015\ {\rm fm/c}
(\as=0.2), \ 0.033\ {\rm fm/c}(\as=0.3)\,.
\end{equation}

The dependence of $t_{\rm th}$ on $\as$ proves to be considerable weaker
in our cascade calculations compared to what is estimated in the \BUp scenario.
In Ref. \cite{XU07} the authors found that the time scale of thermalization
is the inverse of the total transport rate, which is proportional to
$\as^{-2} (\ln \as)^{-2} T^{-1}$ \cite{XGviscos} ($T\sim Q_s$ in the 
present case).
This scaling indeed holds for the thermalization times from our calculations:
\begin{equation}
t_{\rm th}\as^{2}(\ln \as)^2=0.09\ {\rm fm/c}(\as=0.1), 
\ 0.1\ {\rm fm/c}(\as=0.2), \ 0.1\ {\rm fm/c}(\as=0.3)\,.
\end{equation}

Therefore, the time scale of thermalization is of order 
$\as^{-2} (\ln \as)^{-2} Q_s^{-1}$, which is smaller than the \BUp
prediction. The quick thermalization observed here is consistent with
the findings from the previous studies \cite{XU,XU07,XUG06}: The gluon
bremsstrahlung favors large-angle radiation due to the LPM suppression,
which is the reason for the dominance of the pQCD $gg\leftrightarrow ggg$
processes in thermal equilibration.

\begin{figure}[ht]
\begin{center}
\includegraphics[height=7cm]{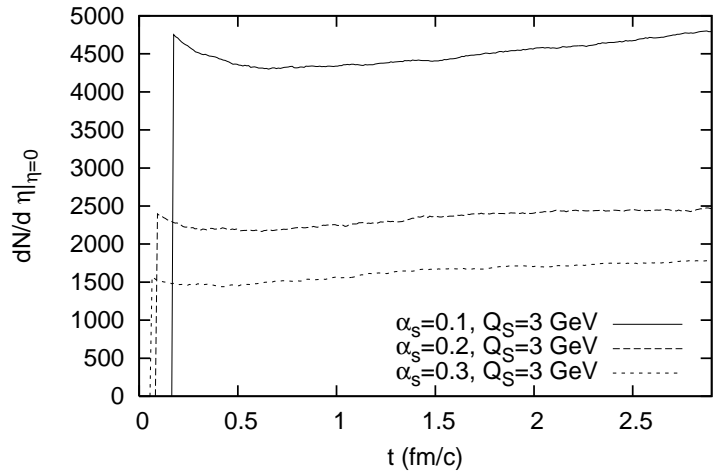}
\end{center}
\caption{Gluon number per space time rapidity in calculations with
fixed $Q_s=3$ GeV and $\as=0.1$, $0.2$, and $0.3$.
}
\label{pnum_small_as}
\end{figure}

Figure \ref{pnum_small_as} shows the time evolution of the total
gluon number per space time rapidity.
The initial condition with $\as=0.2$ and $Q_s=3$ GeV could be appropriate
for Pb-Pb collisions at the maximal energy at the LHC. In this case the gluon number also decreases at the beginning.
Thus, \BUp thermalization might not be favored at LHC.

\begin{figure}[ht]
\begin{center}
\includegraphics[height=7cm]{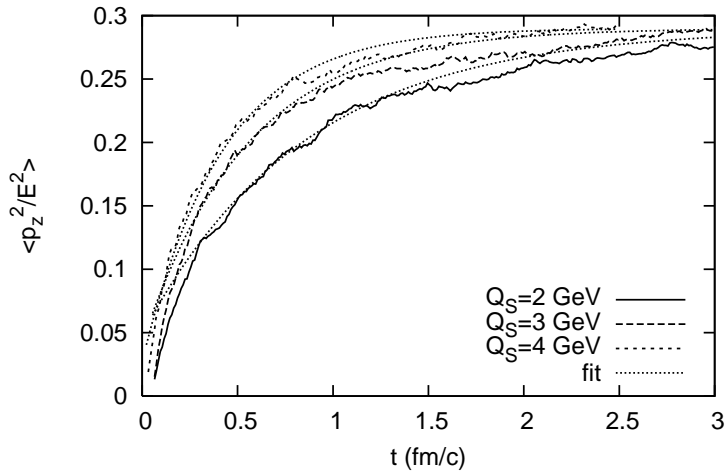}
\end{center}
\caption{Momentum isotropization. CGC initial conditions with
$\as=0.3$ and $Q_s=2$, $3$, and $4$, respectively.
}
\label{psp3}
\end{figure}

Momentum isotropization, which is an important part of kinetic equilibration,
can be described by the time evolution of
$\langle p_z^2/E^2 \rangle$ \cite{XU07} shown in Fig. \ref{psp3}, where CGC initial conditions are used with $\as=0.3$ and $Q_s=2$,
$3$, and $4$, respectively.
$\langle \cdot \rangle$ denotes the average
over gluons in the central space time rapidity. Due to expansion,
$\langle p_z^2/E^2 \rangle$, which is extracted in a finite spatial volume,
saturates at a value that is slightly smaller than its equilibrium value
$1/3$. We fit $\langle p_z^2/E^2 \rangle(t)$ at time $t_0$ using 
a relaxation ansatz
\begin{equation}
\label{relax}
F(t)=\frac{1}{3}+\left ( \langle \frac{p_z^2}{E^2} \rangle (t_0)- 
\frac{1}{3} \right ) \, {\rm exp} \left(-\frac{t-t_0}{\theta_{\rm rel}(t)}
\right )
\end{equation}
where $\theta_{\rm rel}$ gives the time scale of momentum isotropization
\cite{XU07}. The fits are shown in Fig. \ref{psp3}. Using $t_0=0.5$ fm/c, 
we find $\theta_{\rm rel}=0.85$, $0.52$, and $0.42$ fm/c for 
$Q_s=2$, $3$, and $4$ GeV, respectively. These time scales are smaller
than thermalization time scales [see Eq.(\ref{tau_th})], which indicates
that momentum isotropization is completed before (nearly)
full thermalization with quasi-ideal hydrodynamical expansion.

\subsection{Ratio of the shear viscosity to the entropy density}
\label{sec5_4}
As already noticed in the previous subsection, $e\, t^{4/3}$ increases
slightly at later times [see Fig. \ref{dEtdy}(b)], which shows
a deviation from ideal hydrodynamics. However, the agreements of
$T\, t^{1/3}$ and $dE_T/dy \cdot t^{1/3}$ with ideal hydrodynamics
[see Figs. \ref{temp} and \ref{dEtdy}(a)] indicates that the
shear viscosity (or better the ratio of the shear viscosity to the
entropy density) is small.

In the Navier-Stokes approximation, the diagonal elements of the stress
tensor are given \cite{DG85} in the rest frame by
\begin{equation}
T_{ii}=P-2\eta\left(\frac{\partial u_i}{\partial x_i}-
\frac{1}{3}\vec{\nabla} \cdot \vec{u} \right )-
\kappa\vec{\nabla} \cdot \vec{u}
\end{equation}
where $\eta$ denotes the shear viscosity, $\kappa$ the bulk viscosity,
and $P$ the pressure. We then obtain
\begin{eqnarray}
\label{shv}
\eta &=& \frac{T_{xx}+T_{yy}-2\,T_{zz}}{2\,(3\,\partial_z u_z
-\vec \nabla\cdot \vec u)}\,, \\
\label{bv}
\kappa &=& \frac{3\,P-T_{xx}-T_{yy}-T_{zz}}{3\,\vec \nabla\cdot \vec u}\,.
\end{eqnarray}
For the system of massless gluons where $e=3P$, the bulk viscosity
vanishes. The flow velocity $\vec u$ is expected to be approximately the 
same as given in \cite{bjorken}. For an ideal hydrodynamical expansion 
$\vec{u} \approx (0,0,z/t)$ when the gluonic system is (nearly) thermalized.
Thus, we obtain
\begin{equation}
\label{eta}
\eta=\frac{t}{4}\left(T_{xx}+T_{yy}-2 T_{zz}\right)
\end{equation}
where $T_{xx}$, $T_{yy}$ and $T_{zz}$ can be extracted from the
numerical calculations. Note that the divergence of the flow velocity
can be better extracted from the simulations because it relates to
the transport collision rate, as derived in \cite{XGviscos}.

Due to large numerical uncertainties the entropy density it is difficult to extract $s$ from the simulations. 
Therefore, the formula
\begin{equation}
\label{entropy}
s=-\int \frac{d^3p}{(2\pi)^3} \, f_{\rm eq}(p,x) \,[\ln f_{\rm eq}(p,x)-1]
=4n-n\, \ln (\lambda)
\end{equation}
with $f_{\rm eq}(p,x)=\lambda \, d_G \, e^{-\frac{E}{T}}$ is used, which is applied to systems in kinetic equilibrium.
Here $n$ is the gluon density and $\lambda=n/n_{\rm th}$ denotes
the gluon fugacity where $n_{\rm th}=d_G T^3/\pi^2$ is the gluon density
in thermal equilibrium. The entropy density calculated in
Eq. (\ref{entropy}) is, thus, larger than the true value in the simulations.

\begin{figure}[ht]
\begin{center}
\includegraphics[height=7cm]{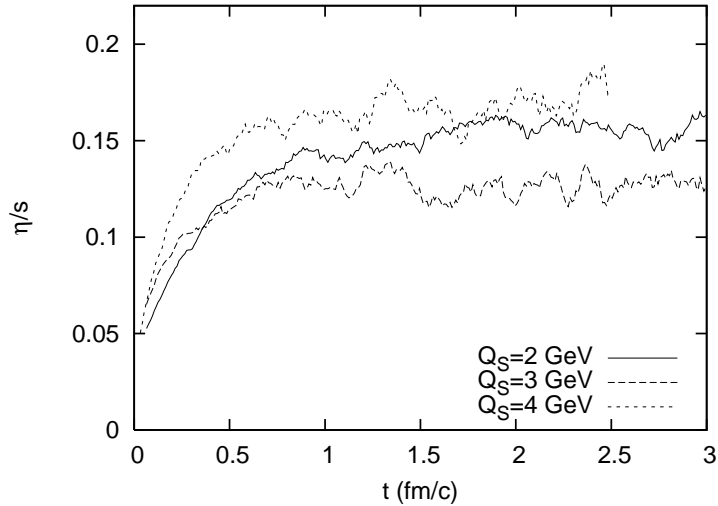}
\end{center}
\caption{Ratio of the shear viscosity to the entropy density.
CGC initial conditions with $\as=0.3$ and $Q_s=2$, $3$, and $4$,
respectively.
}
\label{etaovers}
\end{figure}

Figure \ref{etaovers} shows the $\eta/s$ ratio in the calculations
with $\as=0.3$ and $Q_s=2$, $3$, and $4$, respectively.
Before $0.5$ fm/c the values of $\eta/s$ are not reliable because
at early times the gluon system is far from equilibrium and, thus, 
Eqs. (\ref{eta}) and (\ref{entropy}) are not valid. From $0.5$ fm/c on 
the $\eta/s$ ratio is nearly constant and has a weak dependence
on $Q_s$, $\eta/ s \approx 0.15$, which is exactly the same as that 
obtained in full 3+1 dimensional BAMPS calculations with $\as=0.3$
and minijets type initial conditions for Au+Au collisions at RHIC
energies \cite{XGSv2}. This verifies that the $\eta/s$ ratio determines
the behavior of the late dynamics and, thus, depends only on the coupling
$\as$, but not on initial conditions. Moreover, the $\eta/s$ ratio
obtained is small and close to the lower bound from the AdS/CFT
conjecture \cite{ads}. The smallness of the $\eta/s$ ratio corresponds to
the efficiency of the pQCD $gg\leftrightarrow ggg$ processes in
thermal equilibration, because the $\eta/s$ ratio is inversely proportional
to the total transport collision rate and the transport
collision rate of $gg\leftrightarrow ggg$ processes is $6-7$ times larger
than that of $gg\to gg$ collisions \cite{XGviscos}.

\section{Conclusion}
\label{con}
Using the parton cascade BAMPS, we have studied the thermalization of potential 
color glass condensates, which might be appropriate for the initial
conditions of high energy heavy ion collisions.
The main emphasis is put on the comparison of the thermalization observed
in our calculations to that in the \BUp scenario. We found that several
aspects of the real thermalization might be different compared to
the \BUp scenario. The difference arises because the back reactions of bremsstrahlung, $3\to 2$ processes, play a significant role. They are
completely absent in the \BUp scenario.

First, the radiation of gluons is hindered by $3\to 2$ processes
according to detailed balance. Therefore, for the idealistic form chosen
for CGC, the total gluon number will increase unless $\as Q_s$ is 
very large, which is not realistic at RHIC and LHC. We showed that
for realistic initial conditions the total gluon number decreases
 early in the expansion. An exorbitant increase in soft gluons, as
predicted in the \BUp scenario, is not possible. Thus, no thermal bath
of soft gluons will be built up.

Second, thermal equilibration of soft and hard gluons occurs
roughly on the same time scale due to $2\to 3$ and $3\to 2$ processes,
respectively. The energy flows into both the soft and hard sectors
at the same time, which is potentially similar to the phenomenon of 
''avalanche'' 
as observed in the field isotropization driven by the plasma instability.
This behavior contradicts the \BUp picture where
soft gluons form a thermal bath and thermalize first whereas hard gluons
loose energy to the thermal bath and, thus, thermalize later.

Finally, the time scale of thermalization is determined for various values
of $\as$ and $Q_s$. It spreads from $0.55$ fm/c to $1.75$ fm/c for
$Q_s=2-4$ GeV and $\as=0.1-0.3$.
In agreement with the \BUp scenario, the thermalization
time proves to be proportional to $Q_s^{-1}$, however, its proportionality
to $\as^{-13/5}$ is not seen, but is much weaker: 
$\tau_{\rm th}\sim (\as \ln \as )^{-2} Q_s^{-1}$.

After being thermalized the gluon system shows quasi-hydrodynamical
behavior: The
cooling due to expansion is observed in the steepening of the transverse
momentum spectra. To see how viscous the system is, we extracted the ratio
of the shear viscosity to the entropy density and obtained
$\eta/s \approx 0.15$ for $\as=0.3$. The $\eta/s$ ratio has a weak dependence on $Q_s$
and is close to the lower bound from the AdS/CFT conjecture.
Thus, the considered gluon system acts as being strongly coupled.

The quick thermalization and the smallness of the $\eta/s$ ratio observed
in the present calculations with the CGC initial conditions are
consistent with the findings from the previous 
studies \cite{XU,XU07,XUG06,XGviscos,XGSv2}
using the Glauber-type minijets initial conditions. This demonstrates that
independent of the chosen initial conditions,
the pQCD bremsstrahlung processes (and the back reactions) dominate
the dynamical equilibration and then keep the system behaving like
a nearly perfect fluid. The higher order processes such like $ggg\to ggg$ and $gg\leftrightarrow gggg$ will certainly lead to a larger total transition rate, however, their contributions are suppressed by higher order of $\alpha_s$\cite{Xiong}. Further investigations are needed to quantify these contributions.

The CGC initial conditions presented in this paper are idealistic. 
More realistic initial conditions in high energy nucleus-nucleus collisions can be obtained by considering 
high momentum jet like partons(minijets)\cite{nara}. This will modify the thermalization picture presented in this paper: $ggg\to gg$ interaction may become less dominant at the early stage
of the thermal equilibration, however, it would not affect our conclusion that
thermalization in the hard and soft sectors proceeds on the same time
scale. A study will be presented elsewhere.     

\begin{acknowledgements}
A.E. gratefully acknowledges a fellowship by
the Helmholtz foundation.
\end{acknowledgements}

\end{document}